\documentclass[9pt,conference,table]{IEEEtran}
\usepackage[preprint]{waspaa25}

\usepackage{bm} 

\definecolor{rank1}{RGB}{155, 227, 128}      
\definecolor{rank2}{RGB}{204, 229, 130}
\definecolor{rank3}{RGB}{227, 216, 129}
\definecolor{rank4}{RGB}{222, 179, 124}
\definecolor{rank5}{RGB}{215, 121, 118}      

\title{Towards Reliable Objective Evaluation Metrics for Generative Singing Voice Separation Models}

\name{Paul A. Bereuter$^{1}$,
      Benjamin Stahl$^{1}$,
      Mark D. Plumbley$^{2}$,     
      Alois Sontacchi$^{1}$}
\address{$^{1}$Institute of Electronic Music and Acoustics (IEM), University of Music and Performing Arts, Graz, Austria \\
$^{2}$Centre for Vision, Speech and Signal Processing (CVSSP), University of Surrey, Guildford, UK
}

\begin{document}

\maketitle
\begin{abstract}
Traditional Blind Source Separation Evaluation (BSS-Eval) metrics were originally designed to evaluate linear audio source separation models based on methods such as time-frequency masking. However, recent generative models may introduce nonlinear relationships between the separated and reference signals, limiting the reliability of these metrics for objective evaluation. To address this issue, we conduct a Degradation Category Rating listening test and analyze correlations between the obtained degradation mean opinion scores (DMOS) and a set of objective audio quality metrics for the task of singing voice separation. We evaluate three state-of-the-art discriminative models and two new competitive generative models. For both discriminative and generative models, intrusive embedding-based metrics show higher correlations with DMOS than conventional intrusive metrics such as BSS-Eval. For discriminative models, the highest correlation is achieved by the MSE computed on Music2Latent embeddings. When it comes to the evaluation of generative models, the strongest correlations are evident for the multi-resolution STFT loss and the MSE calculated on MERT-L12 embeddings, with the latter also providing the most balanced correlation across both model types. Our results highlight the limitations of BSS-Eval metrics for evaluating generative singing voice separation models and emphasize the need for careful selection and validation of alternative evaluation metrics for the task of singing voice separation.
 \end{abstract}

\section{Introduction}
\label{sec:intro}

The field of musical source separation (MSS) has greatly benefited from community-driven challenges, including the Stereo Audio Source Separation Evaluation Campaign \cite{vincent2007}, the Signal Separation challenge \cite{sigsep2018}, and the Music Demixing (MDX) challenge \cite{fabbro2024}. A particularly relevant sub-task within MSS, due to its close relation to speech enhancement, is singing voice separation (SVS), i.e., the separation of vocal tracks from a musical mixture. Due to the lack of dedicated metrics, SVS performance is often assessed using BSS-Eval's signal to distortion, signal to interference, signal to artifact and source image to spatial distortion ratio metrics (SDR, SIR, SAR, and ISR), which decompose the estimation error into components via projections onto FIR-filtered subspaces of the target and distorting sources \cite{vincent2006}.

Recently, generative models have been applied to a wide range of audio tasks, including text-to-speech and text-to-audio generation \cite{liu2023audioldm}, as well as tasks traditionally addressed with discriminative models, such as speech enhancement \cite{richter2023speech, soumi2019}, musical source separation \cite{mariani2024multisource,Karchkhadze2025} and singing voice separation \cite{schufo2023, tan2018}. However, for generative models, the BSS-Eval metrics may be unreliable, as the relationship between the separated and reference signals can be nonlinear. While large datasets of human quality ratings have enabled robust evaluations of both discriminative and generative models in the speech domain, similar evaluation methods of generative SVS models remain underexplored.
\subsection{Intrusive Audio Quality Metrics}
Audio quality metrics are typically categorized as \textit{intrusive}, \textit{non-intrusive}, and \textit{non-matching reference}. Intrusive metrics require a reference signal to compute distances or ratios between it and the separated signal. Examples include BSS-Eval and PEASS \cite{peass_toolkit2012}, the latter using regression to approximate subjective ratings. ViSQOL \cite{hines2012} similarly employs a perceptual model with a fitted mapping. For generative tasks such as separation or coding, phase-insensitive metrics such as multi-resolution STFT loss ($\mathcal{L}_{\text{MR}}$) \cite{ryuichi2020} or ViSQOL are more appropriate, as they operate on spectrogram features. Distance measures between reference and test embeddings from speech and audio foundation models using self-supervised learning (SSL), e.g., Contrastive Language-Audio Pretraining (CLAP) \cite{lclap2023}, Acoustic Music Understanding Model with Large-Scale Self-supervised Training (MERT) \cite{li2024mert} or Consistency Autoencoders for Latent Audio Compression (Music2Latent) \cite{pasini2024music2latent}, may also serve as perceptual proxies. For the evaluation of generative models in scenarios where no ground-truth references are available, non-matching reference-based methods are relevant. These approaches entail the comparison of test signals against a pool of non-aligned but semantically similar (non-matching) reference signals in an embedding space e.g. using the Fréchet Audio Distance (FAD) \cite{kilgour19_interspeech, gui2024adapting}, where the Fréchet distance between multivariate Gaussian distributions, fitted to the reference and test embeddings, is calculated.
\subsection{Non-intrusive Audio Quality Metrics}
Non-intrusive methods enable a referenceless perceptual quality estimation, based solely on the separated signal. In the speech domain, XLS-R-SQA \cite{tamm2023} has shown to be a non-intrusive metric that generalizes well on unseen datasets \cite{stahl2025}. A recent universal quality assessment model, Audiobox-Aesthetics \cite{tjandra2025aes}, exhibits strong correlation with human judgements across diverse audio domains. Another universal quality assessment model prompts audio-language models for audio quality assessment (PAM) \cite{deshmukh2024pam}, enabling the identification of quality anchors in CLAP’s embedding space to assess signal quality. For singing voice evaluation, SingMOS \cite{tang2024singmos}, a wav2vec 2.0-based model trained on rated examples from singing voice conversion and coding, is currently the only specialized non-intrusive quality metric. 
\subsection{Contributions and Structure of the Paper}
Despite these developments, the alignment between objective metrics and human perception in SVS remains unclear, particularly in the context of generative models. The study described in this paper investigates the correlation between state-of-the-art intrusive and non-intrusive objective metrics and degradation mean opinion scores (DMOS) stemming from a perceptual evaluation. In contrast to earlier work, which has examined such correlations focusing on tasks like singing voice conversion and coding \cite{tang2024singmos}, this paper specifically focuses on SVS, using stimuli from both discriminative models and novel competitive generative models adapted from recent speech enhancement approaches. By means of this investigation, we aim to identify differences in the performance of audio quality metrics across these model types, potentially informing the selection of suitable evaluation metrics for SVS. 
The remainder of the paper is structured as follows: \Cref{sec:discriminative} details the SVS models, \cref{sec:eval} describes the perceptual evaluation, \cref{sec:obj_eval_and_corr} presents the correlation analysis, and \cref{sec:conclusion} concludes with implications for future SVS evaluation.

\section{Singing Voice Separation Models}
\label{sec:discriminative}
We include three state-of-the-art discriminative mask-based models and two new generative approaches adapted from the field of speech enhancement in our study.
\subsection{Discriminative Models}\label{sec:disc_models}
The Hybrid-Transformer-Demucs (HTDemucs) \cite{rouard_htdemcus_2023}, is chosen as the first discriminative model. HTDemucs is a revised version of the model that won the 2021 Music Demixing Challenge \cite{mitsufuji2021}. It replaces the recurrent bottleneck layer of the initial model \cite{defossez2021hybrid} with a Transformer. We increased the size of HTDemucs by setting the number of transformer channels (\texttt{bottom\_channels}) to 768.
The other two discriminative models included in our investigation are Mel-RoFormer models \cite{bsroformer_LU_2024, wang_melroformer_2024}. The Mel-RoFormer adds a mel-projection module at the beginning of the band-split rope-transformer proposed in \cite{bsroformer_LU_2024} and is amongst the models currently producing the best SDR metrics for singing voice separation \cite{wang_melroformer_2024}. Two variants of the Mel-RoFormer are included: a large pre-trained model MelRoFo\,(L) \cite{jensen2024melbandroformer}, and a scaled down version MelRoFo\,(S) which we trained from scratch. To make MelRoFo\,(S) comparable to HTDemucs in terms of number of parameters, we used the MelRoFo\,(L) architecture from \cite{jensen2024melbandroformer}, and reduced the number of latent features of the RoFormer blocks from $\texttt{dim}=384$ to $192$, but we increased the number of RoPE Transformer encoders from $\texttt{depth}=6$ to $9$. 
\subsection{Generative Models}\label{sec:gen_models}
To incorporate generative singing voice separation models into our study, we looked at recent advances in the field of speech enhancement. Richter et al. \cite{richter2023speech} demonstrate the capability of diffusion-based models for removing noise and reverb in speech signals, by adapting the forward process of the score-based generative model, and starting the diffusion process from the noisy mixture. To adapt this method to singing voice separation, we replaced the noisy and target speech with musical mixtures and target vocals, respectively. We used the NCSNPP score-model architecture from \cite{richter2024ears}, changed the sampling frequency to \SI{44.1}{\kilo\hertz} and the STFT parameters to a window-size of 2046 and a hop-size of 512. We refer to this as the score-based generative model for singing voice separation (SGMSVS). 
The second generative approach included in this study is inspired by the usage of vocoder-based models in speech enhancement to make up for mask-based source separation artifacts and interference residual \cite{soumi2019}. We employ MelRoFo\,(S) to carry out an initial mask-based singing voice separation and then fine-tune the BigVGAN vocoder \cite{lee2023bigvgan} to make up for artifacts introduced by MelRoFo\,(S) (MelRoFo\,(S) + BigVGAN).
\subsection{Used Datasets}
For fair comparison we trained the models HTDemucs, MelRoFo\,(S) and SGMSVS from scratch using the training data of MUSDB18-HQ \cite{musdb18} and MoisesDB \cite{moisesdb23}, which we also used for fine-tuning BigVGAN. To include a strong discriminative baseline in this study, we chose the pre-trained MelRoFo\,(L) checkpoint from \cite{jensen2024melbandroformer}, which was trained on an undisclosed dataset larger than MUSDB18-HQ. The duration of the training samples was \SI{5}{\second}. We applied random mixing, a random gain between 0.25 and 1.25 and channel swapping as data augmentation. Additionally, we enforced that 50\% of all the training data presented to the models were mixtures comprising all four stems (vocals, drums, bass and other). The testing of all models was carried out on \SI{5}{\second} long excerpts of each song of the MUSDB18-HQ test-set (50 test songs in total). These excerpts were selected at random until one was found where the target's RMS value exceeded a threshold of \SI{-30}{\decibel}. The testing and all subsequent evaluations were performed on single-channel audio.
\subsection{Training Details}
\paragraph*{Discriminative Model Training}
HTDemucs was trained using the L1 waveform loss with a learning rate of $10^{-4}$. MelRoFo\,(S) was trained using the loss function proposed in \cite{wang_melroformer_2024} with the training settings defined for MelRoFo\,(L) in \cite{jensen2024melbandroformer}. HTDemucs and MelRoFo\,(S) were trained on stereo signals. HTDemucs was trained using a batch-size of 6 and MelRoFo\,(S) with a batch-size of 1.
\paragraph*{Generative Model Training}
For the SGMSVS model we adopted the score matching training setup of \cite{richter2023speech}. To adapt the transformation of the complex-valued STFT inputs from speech to singing voice signals, we empirically set the spectral scaling factor to $\alpha = 0.0516$ and the spectral compression exponent to $\beta = 0.334$. We also found that using $N=45$ diffusion steps and $PC_{\text{step}}=2$ corrector steps increased the SDR performance of SGMSVS. For the model combining MelRoFo\,(S) and BigVGAN, we used the largest model of \cite{lee2023bigvgan} denoted as $\text{bigvgan\_v2\_44khz\_128band\_512x}$ and fine-tuned it for singing voice separation for 650,000 optimisation steps. Due to extensive GPU memory requirements, the generative models were trained on single-channel audio with a batch-size of 1. 
\paragraph*{Model Selection}
We trained all models but HTDemucs for 550 epochs and selected the checkpoint achieving the best SDR (HTDemucs was trained for 590 epochs to ensure convergence). For the vocoder-based model we used the multi-resolution STFT loss from \cref{sec:intr_obj_eval_imp} as the criterion for model selection. 

The number of model parameters, multiply-accumulate operations per second of audio, used GPUs and the approximate amount of training time for all models are summarized in \cref{tab:model_params}.
\begin{table}[t]
\caption{Number of parameters (Params), multiply and accumulate operations per second (MACs/s) used GPUs and training time of each singing voice separation model.}
\label{tab:model_params}
\resizebox{\columnwidth}{!}{%
\centering
\begin{tabular}{l@{\hspace{8pt}}r@{\hspace{8pt}}r@{\hspace{8pt}}r@{\hspace{10pt}}r@{\hspace{8pt}}}
\toprule
 & Params [M] & MACs/s [G]&GPU [RTX]&train. time [d]  \\
\midrule
HTDemucs                & 81.74 &33.21&1$\times$ 4090&4.2 \\
MelRoFo\,(L)    & 228.20&164.22 & unknown & unknown\\
MelRoFo\,(S)           & 70.48 & 70.09&1$\times$ 4090&11.2 \\
MelRoFo\,(S) + BigVGAN & 192.66&243.04&3$\times$ 4090&(fine-tune) 4.7  \\
SGMSVS                  & 64.74 &373&3$\times$ 3090& 11.5 \\
\bottomrule
\end{tabular}
}
\end{table}
\section{Acquisition of Perceptual Audio Quality Ratings}
\label{sec:eval}

\subsection{Degradation Category Rating (DCR) Test} \label{sec:dcr}
A reference-based perceptual evaluation was chosen to account for preexisting audio effects and background vocals in MUSDB18-HQ and MoisesDB, and to explicitly evaluate the preservation of both singer identity and melody. In order to limit the time necessary to rate the separation quality in terms of artifacts and interferences we chose to look for test paradigms defined in the ITU P.808 standard \cite{ITU-P808-2021}, which are commonly used to evaluate perceived speech quality. Such a reference-based P.808 test paradigm was used in \cite{naderi2020an}. There, a comparison category rating (CCR) test was carried out. In \cite{naderi2020an}, participants sequentially listen to two stimuli and are not aware whether the first or second stimuli is the reference. However, due to the fact that the reference signals exhibit preexisting audio effects, which should not be mistaken as artifacts introduced by SVS models, we found it necessary to make participants aware of the reference.  Thus, we chose to carry out a degradation category rating (DCR) test, in which the reference and stimulus under test are presented sequentially in order. The separated signal is to be rated on a discrete five-point scale, where 5 corresponds to ``degradation is inaudible" and 1 to ``degradation is very annoying". The test design and evaluation including the outlier detection were carried out in compliance with the ITU P.808 standard. The test was implemented as an online listening test building on the webMUSHRA framework \cite{schoeffler2018webmushra}. We normalized the loudness of each stimulus to \SI{-18}{LUFS} according to EBU R128 \cite{EBU-R128}. All processed signals of all models lead to a total of 250 test stimuli, which were randomly divided into three groups (two groups comprise 83 and one group 84 audio samples). To each group we added five randomly chosen reference/reference pairs as gold standard questions.

\subsection{DCR Test Results}\label{sec:dcr_results}
We recruited 30 participants from expert groups within the audio community (university staff, PhD students and audio engineers). 12 ratings per audio sample were obtained (three participants completed all three test groups). All participants were asked to rate their own musical experience. Nine participants described themselves as hobbyists, while 21 participants regarded themselves as experts, as they rely on their musical experience in their professional work. In addition to the standardized P.808 data screening procedure to detect outliers for crowd-sourced listening tests \cite{ITU-P808-2021}, participants' responses to the gold standard questions were used to assess their reliability. No participant violated P.808's outlier criteria. Further, no participant rated the gold standard samples below 4 (``degradation is audible but not annoying") more than three times. Thus, all 12 ratings per samples were averaged to compute the DMOS.

\begin{figure}[t]
  \centering
  \centerline{\includegraphics[width=1.0\columnwidth]{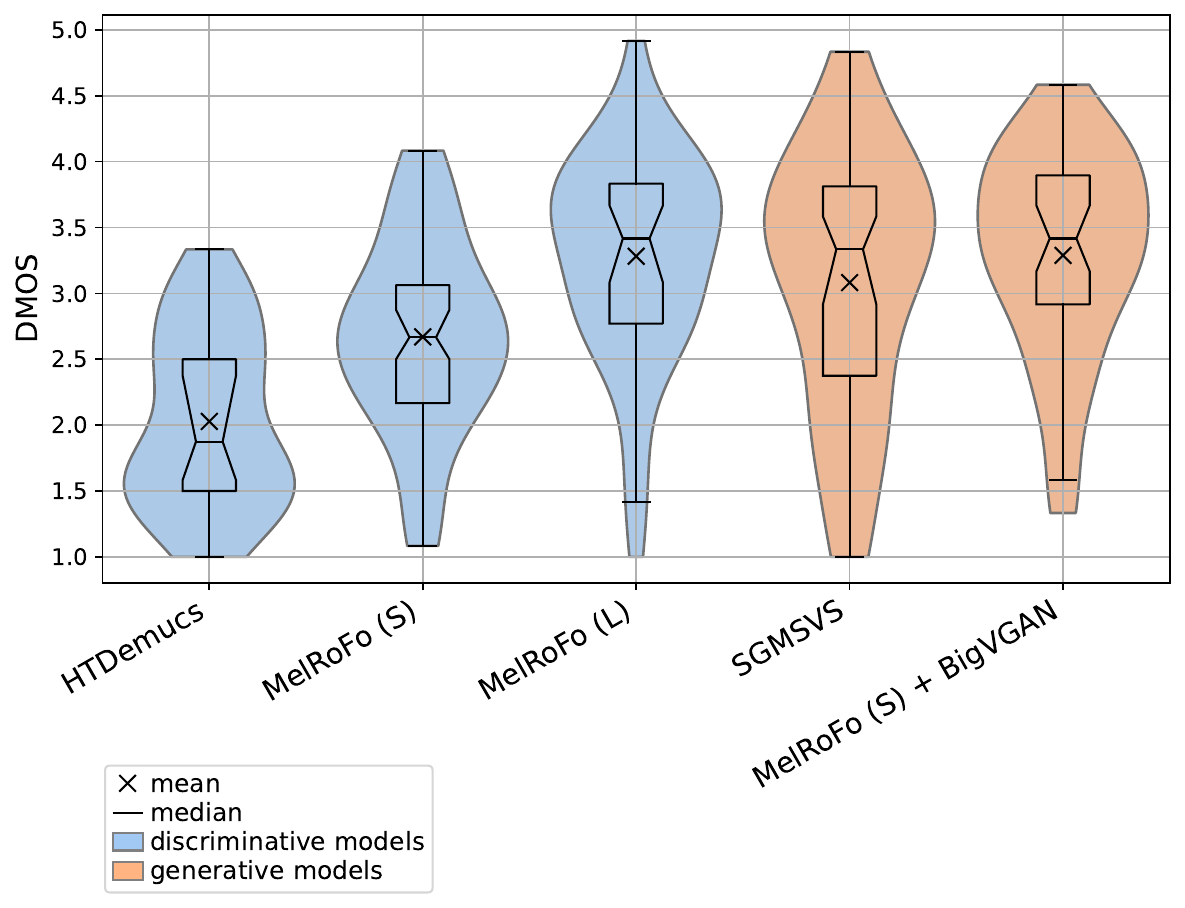}}
  \vspace*{-2mm}
  \caption{Violin and box plots of degradation mean opinion scores (DMOS) grouped by model types. Boxes represent interquartile ranges, and notches indicate bootstrapped confidence intervals of the median.}
  \label{fig:rating_results}
\end{figure}

 The DMOS distributions for the investigated SVS models are visualized in \cref{fig:rating_results}. 
The lowest-scoring model is HTDemucs followed by MelRoFo\,(S). The three best-performing models are SGMSVS, MelRoFo\,(L) and MelRoFo\,(S) + BigVGAN, with relatively small differences in mean and median DMOS, showing that the proposed generative models can compete with the discriminative baseline models. 
The model using the fine-tuned BigVGAN vocoder achieves the highest mean DMOS, clearly outperforming the masking-based Mel-RoFo (S) and enhancing perceived separation quality.

\section{Objective Evaluation Metrics and their correlation with DMOS}\label{sec:obj_eval_and_corr}
This section details the computation of the intrusive and non-intrusive audio quality metrics and their correlation with the DMOS.
\subsection{Computation of Intrusive Audio Quality Metrics} \label{sec:intr_obj_eval_imp}
To calculate BSS-Eval metrics SDR and SI-SDR we used the \textit{TorchMetrics} framework \cite{Detlefsen2022}. The SAR was calculated using \textit{fast\_bss\_eval} \cite{scheibler2022}.
For calculating BSS-Eval's SIR and ISR, and PEASS' overall, target-related, interference-related, and artifacts-related perceptual score (OPS, TPS, IPS, APS) we used the \textit{PEASS Matlab toolkit (v2.0.1)} from \cite{peass_toolkit2012} and the Python Matlab engine.
The ViSQOL v3 command-line tool proposed in \cite{chinen2020} was used to calculate the audio version of ViSQOL. To do so we had to upsample the audio data to a sampling frequency of \SI{48}{\kilo\hertz}. Regarding the multi-resolution STFT loss ($\mathcal{L}_{\text{MR}}$) we used the auraloss toolbox \cite{steinmetz2020auraloss}. There, $\mathcal{L}_{\text{MR}}$ is calculated as an average of $J$ STFT resolutions, where the loss $\mathcal{L}_j$ for the $j^{\text{th}}$ STFT resolution is calculated as:
\begin{align}
    \mathcal{L}_j=\frac{||\text{ }|\mathbf{X}_j|-|\hat{\mathbf{X}}_j|\text{ }||_F}{||\mathbf{X}_j||_F}+\frac{1}{M}||\text{log}\left(|\mathbf{X}_j|\right)-\text{log}\left(|\mathbf{X}_j|\right)||_1
\end{align}
where $\mathbf{X}_j:=\langle X_{m,k}\rangle$ denotes the STFT of the target signal and $\hat{\mathbf{X}}_j:=\langle \hat{X}_{m,k}\rangle$ the STFT of the separated signal for the $j^{\text{th}}$ STFT resolution in $j\in\{256,512,1024,2048,4096\}$ with an overlap of 75\%. $M$ denotes the number of time frames, $||\cdot||_F$ the Frobenius and $||\cdot||_1$ the L1 norm. Both $\mathbf{X}_j$ and $\hat{\mathbf{X}}_j$ were perceptually weighted using an A-weighting filter before loss computation. To calculate the intrusive metrics relying on latent embeddings (FAD and MSE) we employed and adapted the code of the FAD-toolkit published with \cite{gui2024adapting}. The pre-trained encoders included in the FAD-toolkit were used to compute the embeddings L-CLAP\textsubscript{aud} ($\text{CL}_\text{a}$) and L-CLAP-\textsubscript{mus} ($\text{CL}_\text{m}$) from \cite{lclap2023}, as well as the $12^{\text{th}}$ layer embeddings of MERT (M-L12) \cite{li2024mert}. In addition, Music2Latent embeddings (M2L) \cite{pasini2024music2latent} were also used to calculate the FAD and MSE measures. Similar to the per-song FAD used for outlier detection in \cite{gui2024adapting}, we calculate an intrusive variant of the Fréchet audio distance as
\begin{align}
\text{FAD}_{\text{song2song}}=||\mathbf{\mu}-\hat{\mathbf{\mu}}||^2 + \text{trace}\left(\mathbf{\Sigma}+\hat{\mathbf{\Sigma}}-2\sqrt{\mathbf{\Sigma}\hat{\mathbf{\Sigma}}}\right)
\end{align}
where $\mathbf{\mu}$ and $\mathbf{\Sigma}$ are mean and covariance matrix of a multivariate normal distribution over the time-resolved target embeddings and $\hat{\mathbf{\mu}}$ and $\hat{\mathbf{\Sigma}}$ parameterize the multivariate distribution for time-resolved embeddings of the separated signal. To include a metric which accounts for the temporal order of the time-resolved embeddings, the mean squared error (MSE) between the embeddings of the reference and separated signal is computed. 
\subsection{Computation of Non-Intrusive Audio Quality Metrics} \label{sec:nonintr_obj_eval_imp}
Five non-intrusive metrics (PAM, SingMOS, XLS-R-SQA, PQ and CU) were investigated. The PAM metric was included and calculated with the resources published at \cite{deshmukh2024pam}. For SingMOS we used the code that accompanies \cite{tang2024singmos} and for XLS-R-SQA we used the openly accessible Python package of \cite{tamm2023}. The production quality (PQ) and content usefulness (CU) metrics of Meta's Audiobox-AES model were calculated using the Python package proposed in \cite{tjandra2025aes}. Audiobox-AES, SingMOS and XLS-R-SQA operate on \SI{16}{\kilo\hertz}, however only for SingMOS and XLS-R-SQA manual downsampling is necessary, as Audiobox-AES automatically resamples each input signal.
\subsection{Correlation between Objective Metrics and DMOS}\label{sec:corr_results}
\begin{figure*}[t]
  \centering
  \centerline{\includegraphics[width=\linewidth]{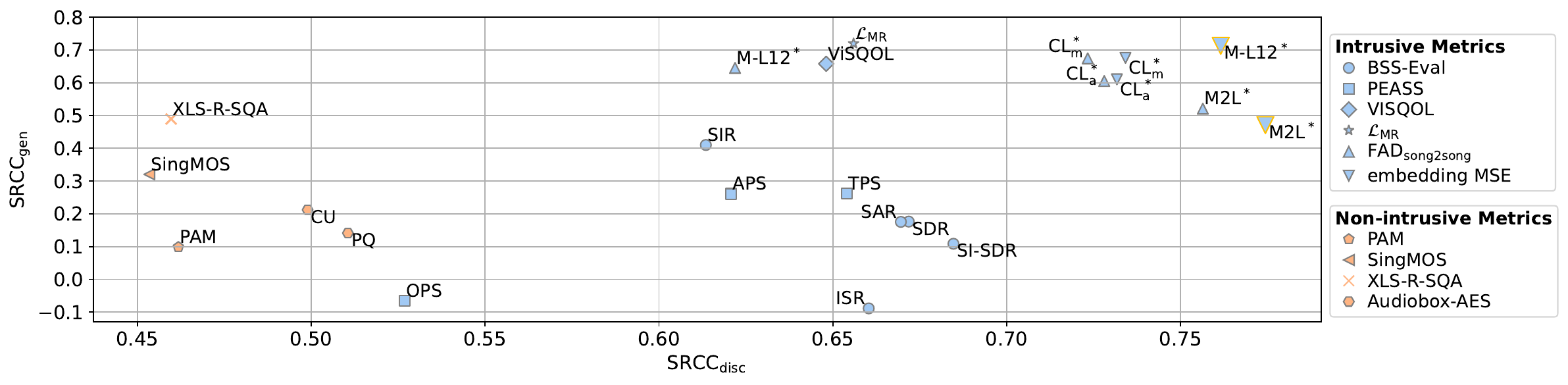}}
  \vspace*{-3mm}
  \caption{Trade-off between Spearman's rank correlation coefficients of objective metrics and DMOS used for evaluation of discriminative ($\text{SRCC}_{\text{disc}}$) and generative models ($\text{SRCC}_{\text{gen}}$). Blue color indicates an intrusive and orange color a non-intrusive metric. All abbreviations are defined in \cref{sec:intro}, \ref{sec:intr_obj_eval_imp} and \ref{sec:nonintr_obj_eval_imp}. Each marker is labeled with the corresponding metric names. Abbreviations marked with $^*$ denote intrusive embedding-based metrics. The metrics exhibiting the highest $\text{SRCC}_{\text{disc}}$ and $\text{SRCC}_{\text{gen}}$ are indicated with larger markers. For consistency, correlations of metrics where lower scores imply better quality (multi-resolution STFT loss, MSE and FAD metrics) have been multiplied by -1.}
  \label{fig:corr_results}
\end{figure*}

The correlations between DMOS and the objective metrics are analyzed separately for both model types (discriminative and generative). Each metric mentioned in \cref{sec:intr_obj_eval_imp} and \ref{sec:nonintr_obj_eval_imp} was evaluated for all separated signals processed by each of the five models. Then 150 DMOS ratings for the discriminative models and all 100 data points of the generative models are utilized to calculate Spearman's rank-based correlation coefficient (SRCC) \cite{spearman1904} for each intrusive and non-intrusive metric. The SRCC, as opposed to Pearson correlation coefficient, was chosen because of its capability to measure monotonicity. The calculated correlation coefficients for all intrusive and non-intrusive metrics are summarized in \cref{fig:corr_results}. The figure illustrates the trade-off between the correlation performance of each metric on the subset of discriminative models ($\text{SRCC}_{\text{disc}}$, shown on the x-axis) and their corresponding SRCC values for the generative model subset ($\text{SRCC}_{\text{gen}}$, shown on the y-axis). Metric types (intrusive/non-intrusive) are highlighted in different colors and each metric is represented by a distinct marker type and their names are annotated. For the $\text{FAD}_{\text{song2song}}$ and MSE metrics, the annotated abbreviations indicate the embeddings used for their calculation, which are marked in \cref{fig:corr_results} with an asterisk (*).

\paragraph*{Intrusive metrics}
\cref{fig:corr_results} shows that, most BSS-Eval and PEASS metrics (SDR, SI-SDR, SAR, SIR, ISR, TPS and APS) exhibit a reasonable correlation when it comes to evaluating discriminative models ($0.6<\text{SRCC}_{\text{disc}}<0.7$). However, they are not representative when it comes to evaluating generative models ($\text{SRCC}_{\text{gen}}<0.5$). PEASS' IPS metric was omitted from \cref{fig:corr_results}, as it shows no significant monotonic correlation with DMOS ratings for either the discriminative or generative subset. The weak correlation between BSS-Eval or PEASS metrics and DMOS for generative models most likely stems from nonlinearities between the separated and target signals, despite possible perceptual similarity. This reduces the reliability of these metrics for evaluating generative models. A better trade-off between $\text{SRCC}_{\text{disc}}$ and $\text{SRCC}_{\text{gen}}$ is evident for ViSQOL and $\mathcal{L}_{\text{MR}}$. Compared to BSS-Eval, ViSQOL and $\mathcal{L}_{\text{MR}}$ exhibit a higher correlation for the evaluation of generative models ($\text{SRCC}_{\text{gen}}\approx 0.7$), while maintaining a similar level of correlation to BSS-Eval for discriminative models.
\paragraph*{Non-intrusive metrics}
When examining the correlation coefficients for the investigated non-intrusive quality metrics, we observe that none of the metrics are able to effectively map perceived source separation degradations to quality scores with a correlation above 0.51 for both model types. The highest correlation for generative model evaluation is achieved by the speech quality metric XLS-R-SQA.
Since XLS-R-SQA has been trained on substantially more data than SingMOS, this indicates that the limited training data available for singing voice, as well as the lack of source separation degradations in existing datasets e.g. the SingMOS training data, restrict the effectiveness of such non-intrusive metrics for evaluating singing voice separation models.

\paragraph*{Embedding-based intrusive metrics}
The best trade-off between the evaluation of discriminative and generative models is achieved by the embedding-based intrusive $\text{FAD}_{\text{song2song}}$ and the embedding MSE. Compared to $\text{FAD}_{\text{song2song}}$, the embedding MSE yields higher $\text{SRCC}_{\text{disc}}$ and comparable $\text{SRCC}_{\text{gen}}$ values across all embeddings, suggesting that accounting for temporal resolution of embeddings better reflects the perceptual evaluation results, at least for discriminative models. The highest $\text{SRCC}_{\text{disc}}$ is obtained using the Music2Latent MSE  ($\text{SRCC}_{\text{disc}}=0.77$). However, its correlation with DMOS for generative models ($\text{SRCC}_{\text{gen}}< 0.5$) remains unreliable. The most balanced correlation across both model types is evident for the MSE calculated on MERT-L12 embeddings.
Please note that all correlation coefficients for FAD and MSE metrics as well as for the multi-resolution STFT loss were multiplied with -1 for consistency. 

\section{Conclusion}\label{sec:conclusion}
The presented results clearly demonstrate the shortcomings of BSS-Eval and PEASS metrics when evaluating generative singing voice separation models. Also, none of the non-intrusive metrics exhibit sufficient correlation with the DMOS data. The discussed embedding-based intrusive metrics, $\text{FAD}_{\text{song2song}}$ and the embedding MSE, have been shown to be useful alternatives, when the right embeddings are chosen. The embedding MSE appears to be the better option for evaluating discriminative singing voice separation models, as it accounts for the temporal resolution of the embeddings. The MERT-L12 embedding MSE achieves the most balanced trade-off overall and seems to serve as a reliable proxy for perceptual ratings when evaluating both discriminative and generative singing voice separation models. In addition, the results of the DCR test show that the proposed generative models can compete with the best discriminative baseline model in our investigation. 

All evaluation audio samples, DMOS scores, computed metrics, and a Python script to calculate $\text{SRCC}_{\text{disc}}$ and $\text{SRCC}_{\text{gen}}$ are available at \url{https://doi.org/10.5281/zenodo.15911723}. Additional resources for training, inference, and evaluation of all five SVS models can be found in the GitHub repository: \url{https://github.com/pablebe/gensvs_eval}. Further details and audio examples are provided on the companion website \url{https://pablebe.github.io/gensvs_eval_companion_page/}.
\section{Acknowledgements}
The majority of this work was funded by the Austrian Federal Ministry of Women, Science and Research (BMFWF) through the Marietta Blau-Grant awarded by Austria's Agency for Education and Internationalisation (OeAD). This work was supported by the Engineering and Physical Sciences Research Council [grant numbers EP/T019751/1, EP/Y028805/1]. For the purpose of open access, the authors have applied a creative commons attribution (CC BY) licence to any author accepted manuscript version arising.
\clearpage


\clearpage
\bibliographystyle{IEEEtran}
\bibliography{refs25}








\end{document}